# The Role of Metadata in Reproducible Computational Research

Jeremy Leipzig[1]; Daniel Nüst[2]; Charles Tapley Hoyt[3]; Stian Soiland-Reyes[4,5];Karthik Ram[6]; Jane Greenberg[1]

[1] Metadata Research Center, Drexel University, College of Computing and Informatics, Philadelphia PA, USA

[2] Institute for Geoinformatics, University of Münster, Münster, Germany

[3] Laboratory of Systems Pharmacology, Harvard Medical School, Boston, USA

[4] eScience Lab, Department of Computer Science, The University of Manchester, Manchester, UK

[5] INDE lab, Informatics Institute, University of Amsterdam, Amsterdam, The Netherlands

[6] Berkeley Institute for Data Science, University of California, Berkeley, USA

## Abstract

Reproducible computational research (RCR) is the keystone of the scientific method for *in silico* analyses, packaging the transformation of raw data to published results. In addition to its role in research integrity, RCR can significantly accelerate evaluation and reuse. This potential and wide support for the FAIR principles have motivated interest in metadata standards supporting RCR. Metadata provides context and provenance to raw data and methods and is essential to both discovery and validation. Despite this shared connection with scientific data, few studies have explicitly described the relationship between metadata and RCR. This article employs a functional content analysis to identify metadata standards that support RCR functions across an analytic stack consisting of input data, tools, notebooks, pipelines, and publications. Our article provides background context, explores gaps, and discovers component trends of embeddedness and methodology weight from which we derive recommendations for future work.

# Contents

# Introduction

Digital technology and computing have transformed the scientific enterprise. As evidence, many scientific workflows have become fully digital, from the problem scoping stage and data collection tasks to analyses, reporting, storage, and preservation. Another key factor includes federal [1] and institutional [2,3] recommendations and mandates to build a sustainable research infrastructure, to support FAIR principles [4], and reproducible computational research (RCR). Metadata has emerged as a crucial component, supporting these advances, with standards supporting the research life-cycle. Reflective of change, there have been many case studies on reproducibility [5], although few studies have systematically examined the role of metadata in supporting RCR. Our aim in this work is to review metadata developments that are directly applicable to RCR, identify gaps, and recommend further steps involving metadata toward building a more robust RCR environment. To lay the groundwork for these recommendations, we first review the RCR and metadata, examine how they relate across different stages of an analysis, and discuss what common trends emerge from this approach.

# Reproducible Computational Research

*Reproducible Research* is an umbrella term that encompasses many forms of scientific quality - from generalizability of underlying scientific truth, exact replication of an experiment with or without communicating intent, to the open sharing of analysis for reuse. Specific to computational facets of scientific research, *Reproducible Computational Research* (RCR)[6] encompasses all aspects of *in silico* analyses, from the propagation of raw data collected from the wet lab, field, or instrumentation, through intermediate data structures, to open code and statistical analysis, and finally publication. Reproducible research points to several underlying concepts of scientific validity – terms that should be unpacked to be understood. Stodden et al. [7] devised a five-level hierarchy of research, classifying it as – reviewable, replicable, confirmable, auditable, and open or reproducible. Whitaker [8] describes an analysis as "reproducible" in the narrow sense that a user can produce identical results provided the data and code from the original, and "generalisable" if it produces similar results when both data is swapped out for similar data ("replicability"), and if underlying code is swapped out with comparable replacements ("robustness") (Figure 1).

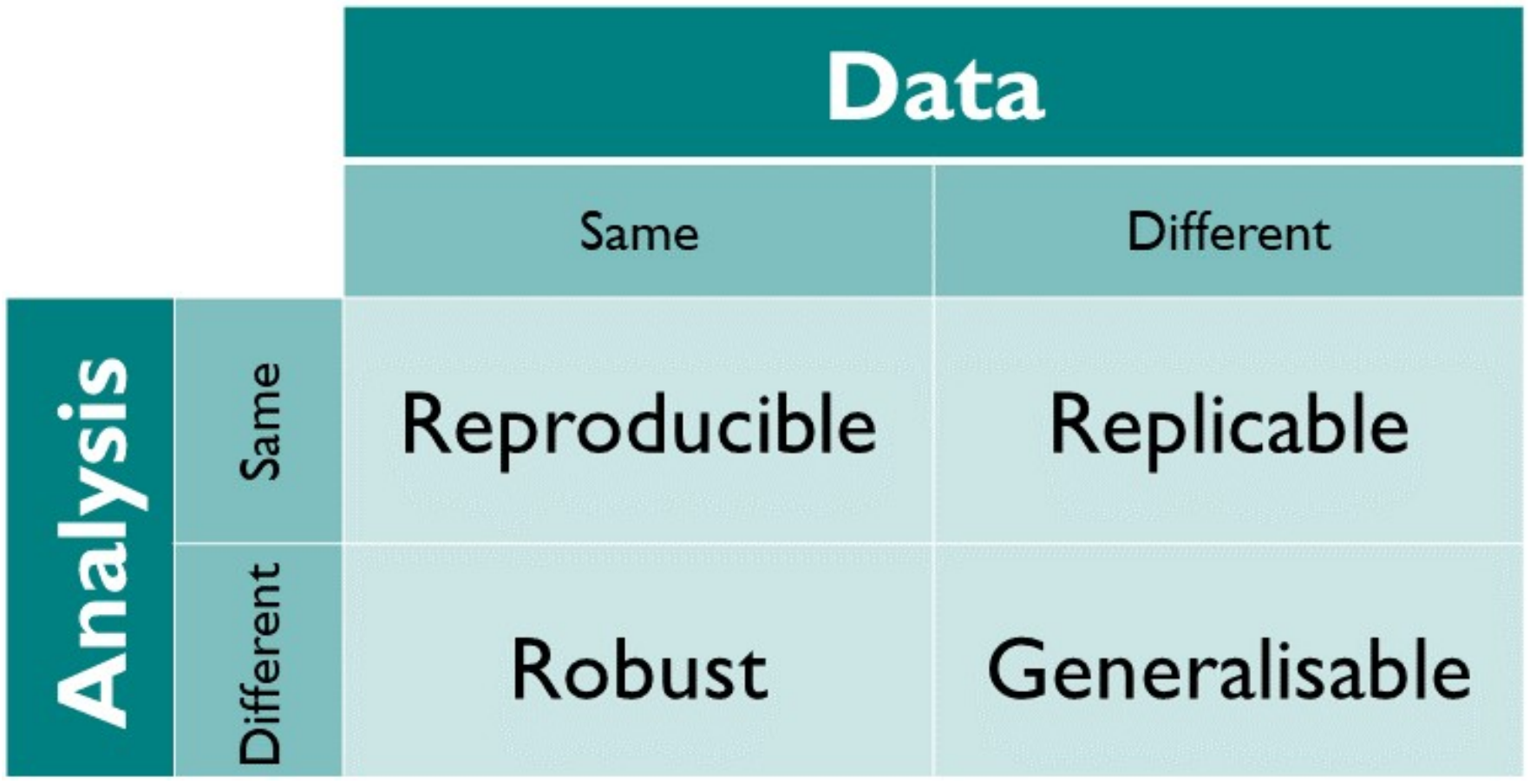

*Figure 1: Whitaker's matrix of reproducibility* [9]

While these terms may confuse those new to reproducibility, a review by Barba disentangles the terminology while providing a historical context of the field [10]. A wider perspective places reproducibility as a first-order benefit of applying FAIR principles: Findability, Accessibility, Interoperability, and Reusability. In the next sections, we will engage reproducibility in the general sense and will use "narrow-sense" to refer to the same data, same code condition.

## Reproducibility Crisis

The scientific community's challenge with irreproducibility in research has been extensively documented [11]. Two events in the life sciences stand as watershed moments in this crisis – the publication of manipulated and falsified predictive cancer therapeutic signatures by a biomedical researcher at Duke and subsequent forensic investigation by Keith Baggerly and David Coombes [12], and a review by scientists at Amgen who could replicate the results of only 6 out of 53 cancer studies [13]. These events involved different aspects - poor data structures and missing protocols, respectively. Together with related studies [14] they underscore recurring reproducibility problems due to a lack of detailed methods, missing controls, and other protocol failures in Inadequate understanding or misuse of statistics, including inappropriate statistical tests and or misinterpretation of results, also plays a recurring role in irreproducibility [15]. Regardless of intent, these activities fall under the umbrella term of "questionable research practices". It bears speculation whether these types of incidents are more likely to occur in novel statistical approaches compared to conventional ones. Subsequent surveys of researchers [11] have identified selective reporting, while theory papers [16] have emphasized the insidious combination of underpowered designs and publication bias, essentially a multiple testing problem on a global scale. We contend that RCR metadata has a role to play in addressing all of these issues and to shift the narrative from a crisis to opportunities [17].

In the wake of this newfound interest in reproducibility, both the variety and volume of related case studies increased after 2015 (Figure 2). Likert-style surveys and high-level publication-based censuses (see Figure 3) in which authors tabulate data or code availability are most prevalent. Additionally, low-level reproductions, in which code is executed, replications in which new data is collected and used, tests of robustness in which new tools or methods are used, and refactors to best practices are also becoming more popular. While the life sciences have generated more than half of these case studies, areas of the social and physical sciences are increasingly the subjects of important reproduction and replication efforts. These case studies have provided the best source of empirical data for understanding reproducibility and will likely continue to be valuable for evaluating the solutions we review in the next sections.

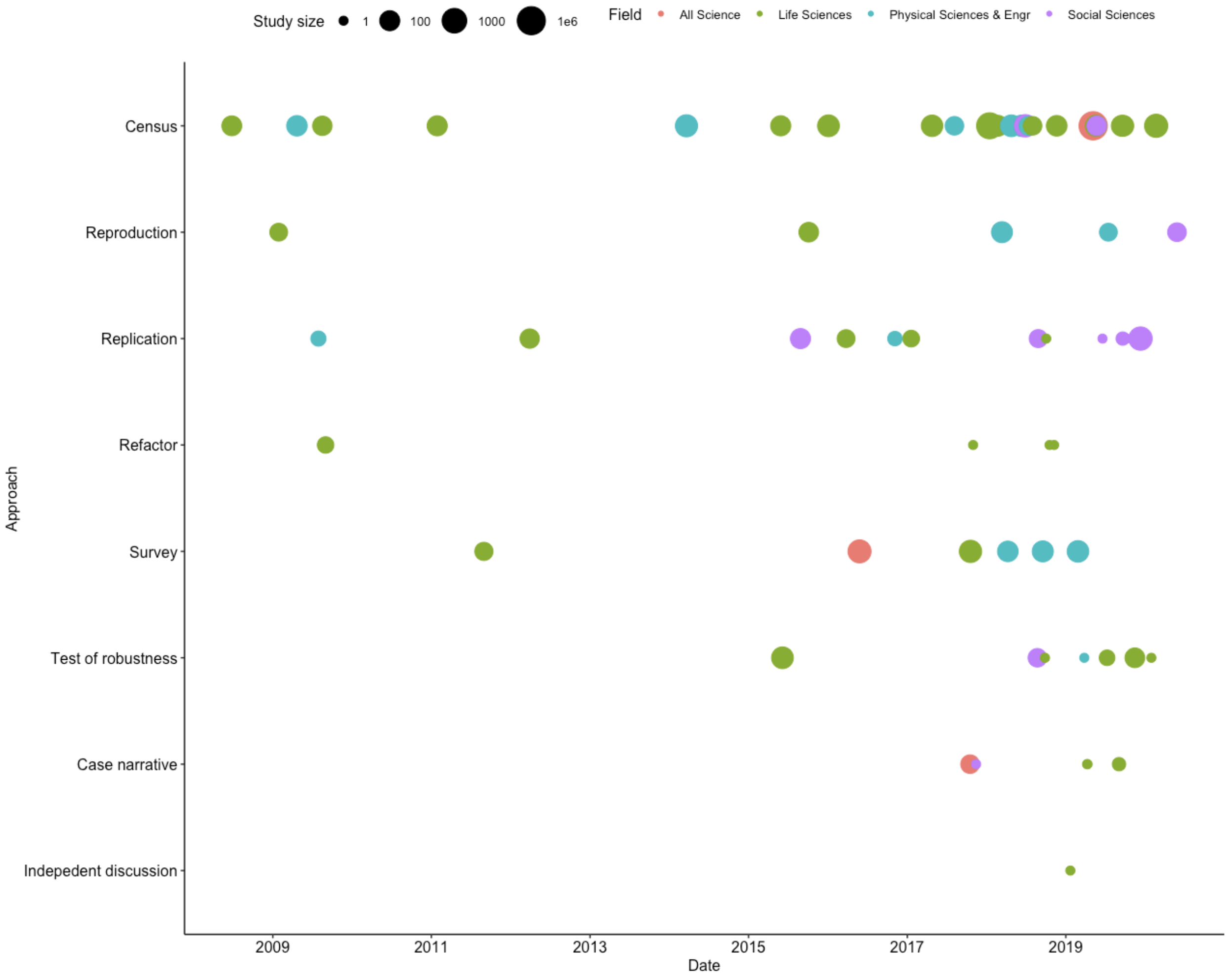

*Figure 2: Case studies in reproducible research [5]. The term "case studies" is used in a general sense to describe any study of reproducibility. A reproduction is an attempt to arrive at comparable results with identical data using computational methods described in a paper. A refactor involves refactoring existing code into frameworks and reproducible best practices while preserving the original data. A replication involves generating new data and applying existing methods to achieve comparable results. A test of robustness applies various protocols, workflows, statistical models, or parameters to a given data set to study their effect on results,*

*either as a follow-up to an existing study or as a "bake-off". A census is a high-level tabulation conducted by a third party. A survey is a questionnaire sent to practitioners. A case narrative is an in-depth first-person account. An independent discussion utilizes a secondary independent author to interpret the results of a study as a means to improve inferential reproducibility.*

## Big Data, Big Science, and Open Data

The inability of third parties to reproduce results is not new to science [18] but the scale of scientific endeavor and the level of data and method reuse suggest replication failures may damage the sustainability of certain disciplines, hence the term "reproducibility crisis." The problem of irreproducibility is compounded by the rise of "big data," in which very large, new, and often unique, disparate or unformatted sources of data have been made accessible for analysis by third parties, and "big science," in which terabyte-scale data sets are generated and analyzed by multi-institutional collaborative research projects. Metadata aspects of big data have been quantitatively studied concerning reuse [19,20], but not reproducibility, despite some evidence big data may play a role in spurious results associated with reporting bias [21]. Big data and big science have increased the demand for high-performance computing, specialized tools, and complex statistics, with attention to the growing popularity and application of machine learning and deep learning (ML/DL) techniques to these data sources. Such techniques typically train models on specific data subsets, and the models, as the end product of these methods, are often "black boxes," i.e. their internal predictors are not explainable (unlike older techniques such as regression) though they provide a good fit for the test data. Properly evaluating and reproducing studies that rely on such algorithms presents new challenges not previously encountered with inferential statistics [22,23]. RCR is typically focused on the last analytic steps of what is often a labor-intensive scientific process that often originates from wet-lab protocols, fieldwork, or instrumentation and these last *in silico* steps present some of the more difficult problems both from technical and behavioral standpoints, because of the amount of entropy introduced by the sheer number of decisions made by an analyst. Developing solutions to make ML/DL workflows transparent, interpretable, and explorable to outsiders, such as peer reviewers, is an active area of research [24].

The ability of third parties to reproduce studies relies on access to the raw data and methods employed by authors. Much to the exasperation of scientists, statisticians, and scientific software developers, the rise of "open data" has not been matched by "open analysis" as evidenced by several case studies [25–28].

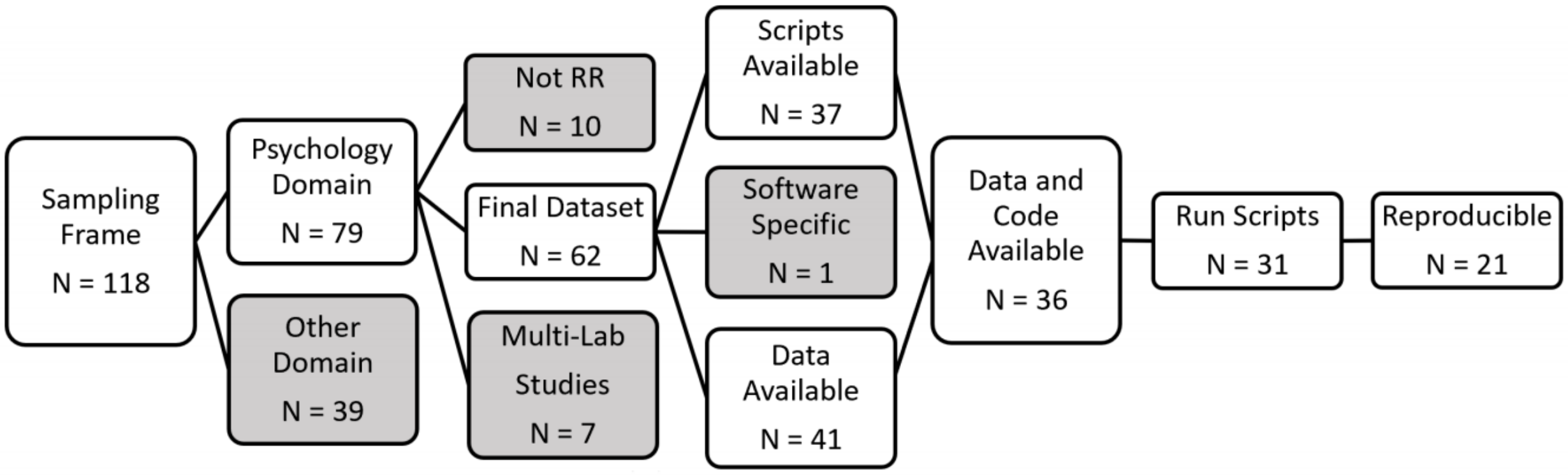

*Figure 3. Censuses like this one by Obels et al. measure data and code availability and reproducibility in this case over a corpus of 118 studies [25].*

Missing data and code can obstruct the peer review process, where proper review requires the authors to put forth the effort necessary to share a reproducible analysis. Software development practices, such as documentation and testing, are not a standard requirement of the doctoral curriculum, the peer-review process, or the funding structure – and as a result, the scientific community suffers from diminished reuse and reproducibility [29]. Sandve et al. [30] identified the most common sources of these oversights in "Ten Simple Rules for Reproducible Computational Research" – lack of workflow frameworks, missing platform and software dependencies, manual data manipulation or forays into web-based steps, lack of versioning, lack of intermediates and plot data, and lack of literate programming or context can derail a reproducible analysis.

An issue distinct from the availability of source code and raw data is the lack of metadata to support reproducible research. We have observed many of the findings from case studies in reproducibility point to missing methods details in an analysis, which can include software-specific elements such as software versions and parameters [31], but also steps along the entire scientific process including data collection and selection strategies, data processing provenance, statistical methods and linking these elements to publication. We find the key concept connecting all of these issues is metadata.

An ensemble of dependency management and containerization tools already exist to accomplish narrow-sense reproducibility [32] – the ability to execute a packaged analysis with little effort from third-party. But context to allow for robustness and replicability, "broad-sense reproducibility," is limited without endorsement and integration of necessary metadata standards that support discovery, execution, and evaluation. Despite the growing availability of open-source tools, training, and better executable notebooks, reproducibility is still challenging [33]. In the following sections, we address these issues, first defining metadata, defining an "analytic stack" to abstract the steps of an in silico analysis, and then identifying and categorizing standards both established and in development to foster reproducibility.

## Metadata

Over the last twenty-five years, metadata has gained acceptance as a key component of research infrastructure design. This trend is defined by numerous initiatives supporting the development and sustainability of hundreds of metadata standards, each with varying characteristics [34,35]. Across these developments, there is a general high-level consensus regarding the following three types of metadata standards [36,37]:

1. *Descriptive metadata*, supporting the discovery and general assessment of a resource (e.g., the format, content, and creator of the resource).
2. *Administrative metadata*, supporting technical and other operational aspects affiliate with resource use. Administrative metadata includes technical, preservation, and rights metadata.
3. *Structural metadata*, supporting the linking among the components of a resource, so it can be fully understood.

There is also general agreement that metadata is a key aspect in supporting FAIR, as demonstrated by the FAIRsharing project (https://fairsharing.org), which divides standards types into "reporting standards" (checklists or templates e.g. MIAME [38]), "terminology artifacts or semantics" (formal taxonomies or ontologies to disambiguate concepts e.g. Gene Ontology [39]), "models and formats" (e.g. FASTA [40]), "metrics" (e.g. FAIRMetrics [41]) and "identifier schemata" (e.g. DOI [42]) [43]. (See Table 1).

*Table 1: Types of FAIRsharing Data and Metadata Standards*

| Type of standard | Purpose |
|---|---|
| Reporting standards | Ensure adequate metadata for reproduction |
| Terminology artifacts or semantics | Concept disambiguation and semantic relationships |
| Models and formats | Interoperability |
| Identifier schemata | Discovery |

Metadata is by definition structured. However, structured intermediates and results that are used as part of scientific analyses and employ encoding languages such as JSON or XML are recognized as primary data, not metadata. While an exhaustive distinction is beyond the scope of this paper, we define RCR metadata broadly as **any structured data that aids reproducibility and that can conform to a standard**. While this definition may seem liberal, we contend that metadata is the "glue" of RCR, and best identified by its function rather than its origins. This general understanding of metadata as a necessary component for research and data management and growing interest in RCR, together with the fact that there are few studies targeting metadata about the analytic stack that motivated the research presented in this paper.

## Goals and Methods

Our overall goal of this work is to review existing metadata standards and new developments that are directly applicable to RCR, identify gaps, discuss common threads among these efforts, and recommend next steps toward building a more robust RCR environment.

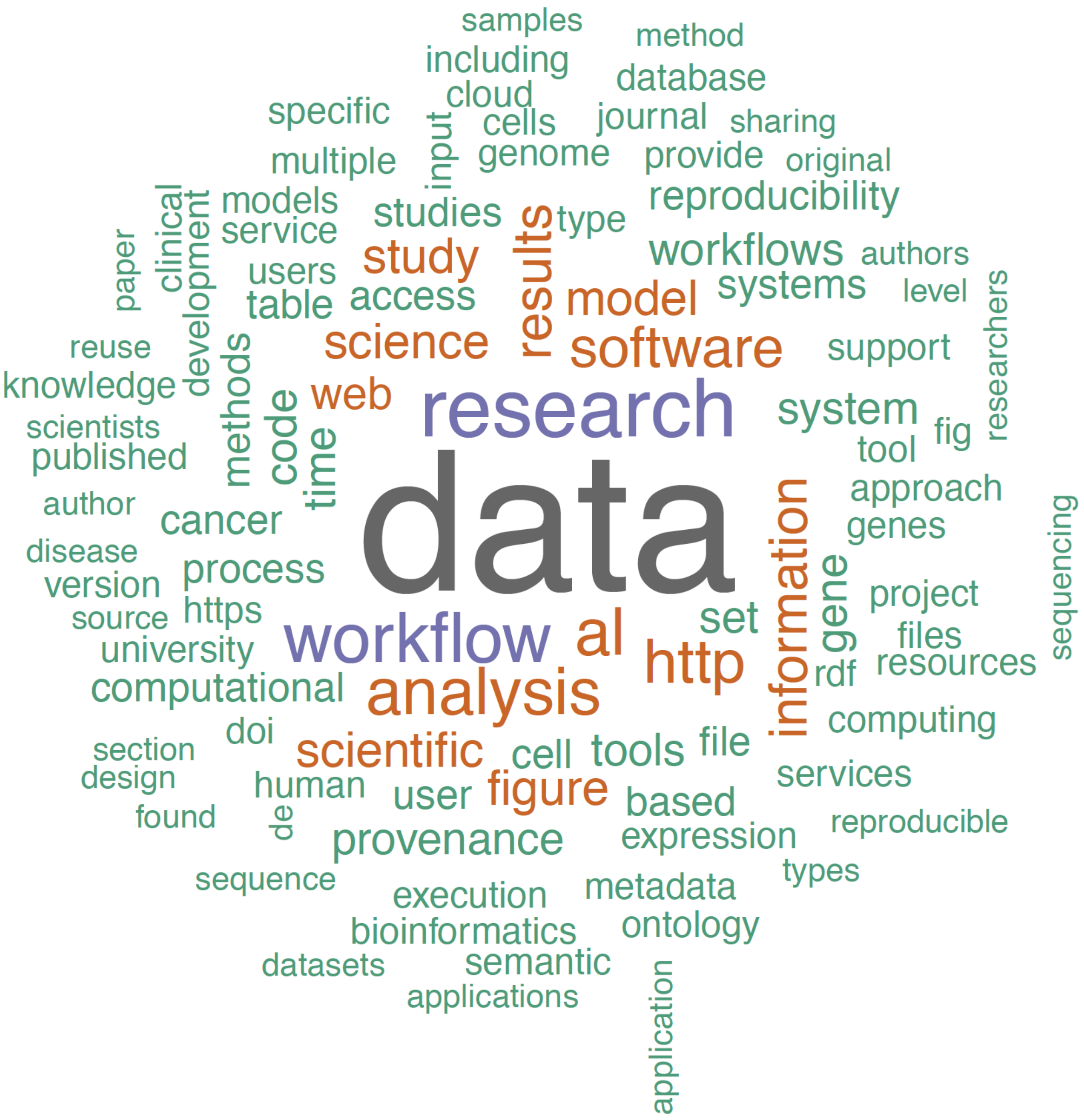

*Figure 4: Terms enriched in the review corpus*

Our method is framed as a state-of-the-art review based on literature (Figure 4) and ongoing software development in the scientific community. Review steps included: **1)** defining key components of the RCR analytic stack, and function that metadata can support, **2)** selecting exemplary metadata standards that address aspects of the identified functions, **3)** assessing the applicability of these standards for supporting RCR functions, and **4)** designing the RCR metadata hierarchy. Our approach was informed, in part, by the Qin LIGO case study [44], catalogs of metadata standards such as FAIRSharing, and comprehensive projects to bind semantic science such as Research Objects [45]. Compilation of core materials was accomplished mainly through literature searches but also perusal of code repositories, ontology catalogs, presentations, and Twitter posts. A "word cloud" of the most used abstract terms in the cited papers reveals most general terms.

### The RCR metadata stack

To define the key aspects of RCR, we have found it useful to break down the typical scientific computational analysis workflow, or "analytic stack," into five levels - 1. input, 2. tools, 3. reports, 4. pipelines, and 5. publication. These levels correspond loosely to the data science (Data understand, prep, modeling, evaluation, deployment), scientific method (formulation, hypothesis, prediction, testing, analysis), and various research lifecycles as proposed by data curation communities (data search, data management, collection, description, analysis, archival, and publication) [46] and software development communities (Plan, Collect, Quality Control, Document, Preserve, Use). However, unlike the steps in the lifecycle we do not emphasize a strong temporal order to these layers, but instead consider them simply interactive components of any scientific output.

## Synthesis Review

In the course of our research, we found most standards, projects, and organizations were intended to address reproducibility issues that corresponded to specific activities in the analytic stack. However, metadata standards were unevenly distributed among the levels. Standards that could arguably be classified or repurposed into two to more areas were placed closest to their original intent.

The synthesis below first presents a high-level summary table, followed by a more detailed description of each of the five levels, specific examples, and a forecast of future directions.

| **Metadata Level** | **Description** | **Examples of Metacontent** | **Examples of Standards** | **Projects and Organizations** |
|---|---|---|---|---|
| 1. Input | Metadata related to raw data and intermediates | Sequencing parameters, instrumentation, | **MIAME**, **EML**, **DICOM** GBIF | OBO, NCBO, FAIRsharing, Allotrope |
| | | spatiotemporal extent | CIF, ThermoML, CellML, DATS, FAANG, ISO/TC 276, NetCDF, OGC, GO | |
| 2.Tools | Metadata related to executable and script tools | Version, dependencies, license, scientific domain | **CRAN DESCRIPTION file, Conda meta.yaml/environment.yml**, **pip requirements.txt**, pipenv Pipfile/Pipfile.lock, Poetry pyproject.toml/poetry.lock, **EDAM**, **CodeMeta**, Biotoolsxsd, DOAP, ontosoft, SWO | Dockstore, Biocontainers |
| 3.Statistical reports and Notebooks | Literate statistical analysis documents in Jupyter or knitr, Overall statistical approach or rationale | Session variables, ML parameters, inline statistical concepts | OBCS, **STATO** SDMX DDI, **MEX**, MLSchema, **MLFlow, Rmd YAML** | Neural Information Processing Systems Foundation |
| 4.Pipelines, Preservation, and Binding | Dependencies and deliverables of the pipeline, provenance | File intermediates, tool versions, deliverables | **CWL, CWLProv, RO-Crate,** RO, WICUS, OPM, PROV-O, ReproZip Config, ProvOne, WES, BagIt, BCO, ERC | GA4GH, ResearchObjects, WholeTale, ReproZip |
| 5.Publication | Research domain, keywords, attribution | Bibliographic, Scientific field, Scientific approach (e.g. "GWAS") | **BEL**, Dublin Core, JATS, ONIX, MeSH, LCSH, MP, Open PHACTS, SWAN, SPAR, PWO, PAV | NeuroLibre, JOSS, ReScience, Manubot |

*Table 2: Metadata standards including: MIAME [38], EML [47], DICOM[48], GBIF [49], CIF [50], ThermoML [51], CellML [52], DATS [53], FAANG [54], ISO/TC 276 [55], GO[39], Biotoolsxsd [56], meta.yaml [57], DOAP [58], ontosoft [59], EDAM [60], SWO [61], OBCS [62], STATO [63], SDMX [64], DDI [65]), MEX [66], MLSchema [67], CWL [68], WICUS [69], OPM [70], PROV-O [71], CWLProv [72], ProvOne [73], PAV [74], BagIt [75], RO [45], RO-Crate [76], BCO [77], Dublin Core [78], JATS [79], ONIX [80], MeSH [81], LCSH [82], MP [83], Open*

*PHACTS [84], BEL [85], SWAN [86], SPAR [87], PWO [88]. Standards in **bold** are featured within this article. Examples of all standards can be found at https://github.com/leipzig/metadata-in-rcr*

## 1. Input

Input refers to raw data from wet lab, field, instrumentation, or public repositories, intermediate processed files, and results from manuscripts. Compared to other layers of the analytic stack, input data garners the majority of metadata standards. Descriptive standards (metadata) enable the documentation, discoverability, and interoperability of scientific research and make it possible to execute and repeat experiments. Descriptive metadata, along with provenance metadata also provides context and history regarding the source, authenticity, and life-cycle of the raw data. These basic standards are usually embodied in the scientific output of tables, lists, and trees which take form in files of innumerable file and database formats as input to reproducible computational analyses, filtering down to visualizations and statistics in published journal articles. Most instrumentation, field measurements, and wet lab protocols can be supported by metadata used for detecting anomalies such as batch effects and sample mix-ups.

While metadata is often recorded from firsthand knowledge of the technician performing an experiment or the operator of an instrument, many forms of input metadata are in fact metrics that can be derived from the underlying data. This fact does not undermine the value of "derivable" metadata in terms of its importance for discovery, evaluation, and reproducibility.

Formal semantic ontologies represent one facet of metadata. The OBO Foundry [89] and NCBI BioPortal serve as catalogues of life science ontologies. The usage of these ontologies appears to follow a steep Pareto distribution, with "Gene Ontology" garnering more than 20,000 term mentions in PubMed, the vast majority of NCBO's 843 ontologies have never been cited or mentioned.

### Examples

In addition to being the oldest, and arguably most visible of RCR metadata standards, input metadata standards serve as a watershed for downstream reproducibility. To understand what input means for RCR, we will examine three well-established examples of metadata standards from different scientific fields. Considering each of these standards reflects different goals and practical constraints of their respective fields, their longevity merits investigating what characteristics they have in common.

#### DICOM - An embedded file header

Digital Imaging and Communications in Medicine (DICOM) is a medical imaging standard introduced in 1985 [90]. DICOM images require extensive technical metadata to support image rendering, and descriptive metadata to support clinical and research needs. These metadata coexist in the DICOM file header, which uses a group/element namespace to designate public restricted standard DICOM tags from private metadata. Extensive standardization of data types, called value representations (VRs) in DICOM, also follow this public/private scheme [91]. The public tags, standardized by the National Electrical Manufacturers Association (NEMA), have served the technical needs of both 2 and 3-dimensional images, as well as multiple frames, and multiple associated DICOM files or "series." Conversely, descriptive metadata has suffered from "tag entropy" in the form of missing, incorrectly filled, non-standard or misused tags by technicians manually entering in metadata [92]. This can pose problems both for clinical workflows as well as efforts to aggregate imaging data for data mining and machine learning. Advanced annotations supporting image segmentation and quantitative analysis have to conform to data structures imposed by the DICOM header format. This has made it necessary for programs such as 3DSlicer [93] and its associated plugins, such as dcqmi [94] to develop solutions such as serializations to accommodate complex or hierarchical metadata.

### EML - Flexible user-centric data documentation

Ecological Metadata Language (EML) is a common language for sharing ecological data [47]. EML was developed in 1997 by the ecology research community and is used for describing data in notable databases, such as the Knowledge Network for Biocomplexity (KNB) repository (https://knb.ecoinformatics.org/) and the Long Term Ecological Network (https://lternet.edu/). The standard enables documentation of important information about who collected the research data, when, and how – describing the methodology down to specific details and providing detailed taxonomic information about the scientific specimen being studied (Figure 5).

```xml
<coverage>
  <geographicCoverage>
    <geographicDescription>Global Oceans</geographicDescription>
    <boundingCoordinates>
      <westBoundingCoordinate>-180</westBoundingCoordinate>
      <eastBoundingCoordinate>180</eastBoundingCoordinate>
      <northBoundingCoordinate>90</northBoundingCoordinate>
      <southBoundingCoordinate>-90</southBoundingCoordinate>
    </boundingCoordinates>
  </geographicCoverage>
  <temporalCoverage>
    <rangeOfDates>
      <beginDate>
        <calendarDate>2008</calendarDate>
      </beginDate>
      <endDate>
        <calendarDate>2013</calendarDate>
      </endDate>
    </rangeOfDates>
  </temporalCoverage>
</coverage>
```

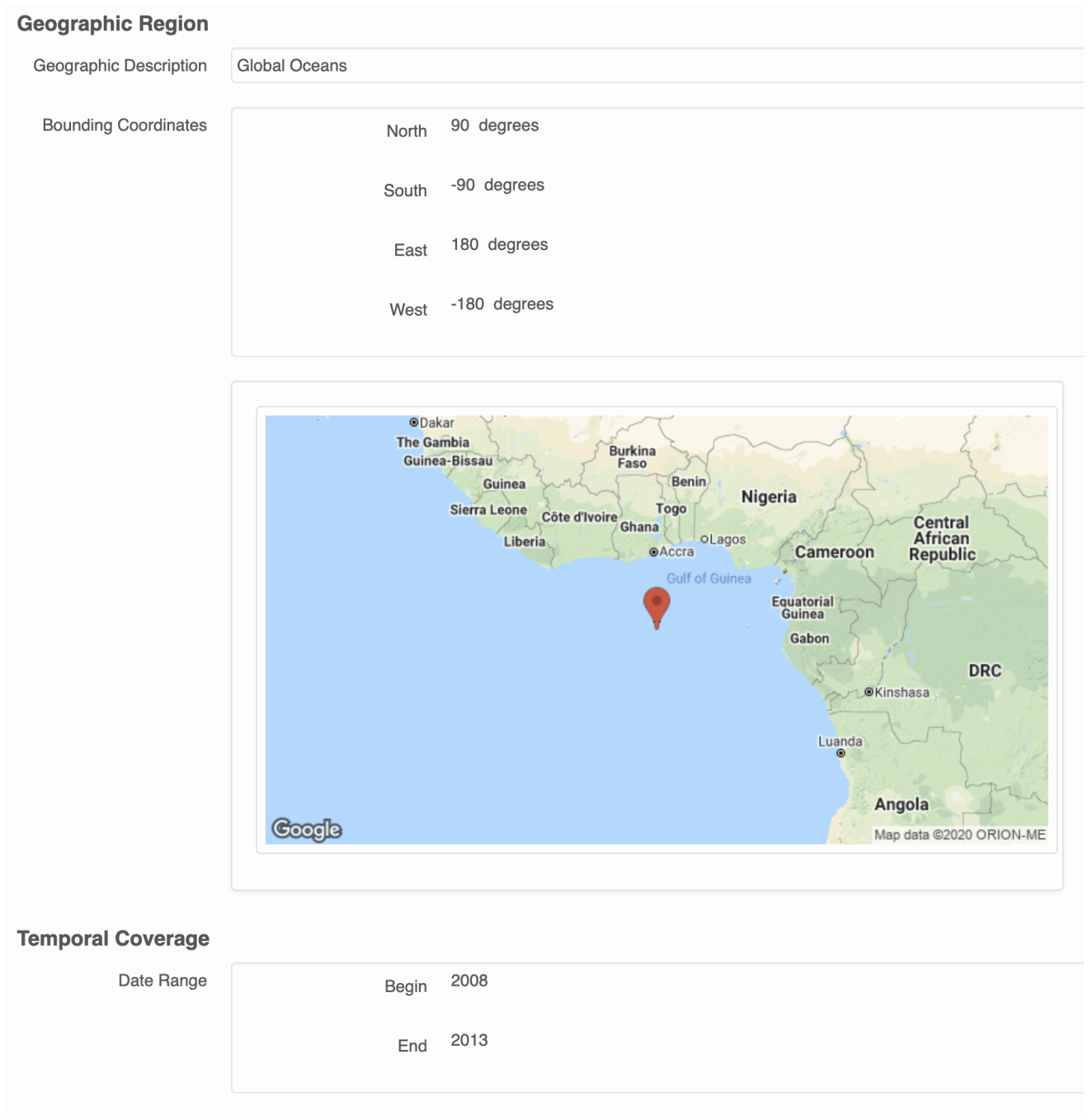

*Figure 5: Geographic and temporal EML metadata and the associated display on Knowledge Network for Biocomplexity (KNB) from Halpern et al.* [95]

### MIAME - A submission-centric minimal standard

Minimum Information About a Microarray Experiment (MIAME) [38] is a set of guidelines developed by the Microarray Gene Expression Data (MGED) society that has been adopted by many journals to support an independent evaluation of results. Introduced in 2001, MIAME allows public access to crucial metadata supporting gene expression data (i.e. quantitative measures of RNA transcripts) via the Gene Expression Omnibus (GEO) database at the National Center for Biotechnology Information and European Bioinformatics Institute (EBI) ArrayExpress. The standard allows microarray experiments encoded in this format to be reanalyzed, supporting a fundamental goal of RCR: to support structured and computable experimental features [96].

```
<?xml version="1.0" encoding="UTF-8" standalone="no"?>

<MINiML
   xmlns="https://www.ncbi.nlm.nih.gov/geo/info/MINiML"
   xmlns:xsi="http://www.w3.org/2001/XMLSchema-instance"
   xsi:schemaLocation="https://www.ncbi.nlm.nih.gov/geo/info/MINiML https://www.ncbi.nlm.nih.gov/geo/info/MINiML.xsd"
   version="0.5.0" >

  <Contributor iid="contrib1">
    <Person><First>Jun</First><Last>Shima</Last></Person>
  </Contributor>

  <Contributor iid="contrib2">
    <Person><First>Fumiko</First><Last>Tanaka</Last></Person>
  </Contributor>

  <Contributor iid="contrib3">
    <Person><First>Akira</First><Last>Ando</Last></Person>
  </Contributor>

  <Contributor iid="contrib4">
    <Person><First>Toshihide</First><Last>Nakamura</Last></Person>
  </Contributor>

  <Contributor iid="contrib5">
    <Person><First>Hiroshi</First><Last>Takagi</Last></Person>
  </Contributor>

  <Database iid="GEO">
    <Name>Gene Expression Omnibus (GEO)</Name>
    <Public-ID>GEO</Public-ID>
    <Organization>NCBI NLM NIH</Organization>
    <Web-Link>https://www.ncbi.nlm.nih.gov/geo</Web-Link>
    <Email>geo@ncbi.nlm.nih.gov</Email>
  </Database>

  <Platform iid="GPL90">
    <Accession database="GEO">GPL90</Accession>
  </Platform>

  <Sample iid="Sample1">
    <Title>
before fermentation
    </Title>
    <Channel-Count>1</Channel-Count>
    <Channel position="1">
      <Source>mRNA T128</Source>
      <Organism>Saccharomyces cerevisiae</Organism>
      <Characteristics>
Typical commercial baker's yeast used in Japan
      </Characteristics>
      <Treatment-Protocol>
```

*Figure 6: An example of MIAME in MINiML format ([https://www.ncbi.nlm.nih.gov/geo/info/MINiML_Affy_example.txt](https://www.ncbi.nlm.nih.gov/geo/info/MINiML_Affy_example.txt))*

MIAME (Figure 6) has been a boon to the practice of meta-analyses and harmonization of microarrays, offering essential array probeset, normalization, and sample metadata that make the over 2 million samples in GEO meaningful and reusable [97]. However, it should be noted that among MIAME and other Investigation/Study/Assay (ISA) standards that have followed suit [98], none offer a controlled vocabulary for describing downstream computational workflows aside from slots to name the normalization procedure applied to what are essentially unitless intensity values.

### Future directions - encoding, findability, granularity

Metadata for input is developing along descriptive, administrative, and structural axes. Scientific computing has continuously and selectively adopted technologies and standards developed for the larger technology sector. Perhaps most salient from a development standpoint is the shift from extensible markup language (XML) to more succinct Javascript Object Notation (JSON) and Yet Another Markup Language (YAML) as preferred formats, along with requisite validation schema standards [99].

The term "semantic web" describes an early vision of the internet based on machine-readable contextual markup and semantically linked data using Uniform Resource Identifier (URI) [100]. Schema.org, a consortium of e-commerce companies developing tags for markup and discovery, such as those recognized by Google Dataset Search [101], has coalesced a stable set of tags that is expanding into scientific domains, demonstrating the potential for findability. Schema.org can be used to identify and distinguish inputs and outputs of analyses in a disambiguated and machine-readable fashion. DATS [53], a Schema.org-compatible tag-suite describes fundamental metadata for datasets akin to that used for journal articles.

Finally, the growing scope for input metadata describing and defining unambiguous lab operations and protocols is important for reproducibility. One example of such an input metadata framework is the Allotrope Data Format, an HDF5 data structure, and accompanying ontology for chemistry protocols used in the pharmaceutical industry [102]. Allotrope uses the W3C Shapes Constraint Language (SHACL) to describe which RDF relationships are valid to describe lab operations.

## 2. Tools

Tool metadata refers to administrative metadata associated with computing environments, compiled executable software, and source code. In scientific workflows, executable and script-based tools are typically used to transform raw data into intermediates that can be analyzed by statistical packages and visualized as, e.g., plots or maps. Scientific software is written for a variety of platforms and operating systems; although Unix/Linux based software is especially common, it is by no means a homogenous landscape. In terms of reproducing and replicating studies, the specification of tools, tool versions, and parameters is paramount. In terms of tests of robustness (same data/different tools) and generalizations (new data/different tools), communicating the function and intent of a tool choice is also important and presents opportunities for metadata. Scientific software is scattered across many repositories in both source and compiled forms. Consistently specifying the location of software using URLs is neither trivial nor sustainable. To this end, a Software Discovery Index was proposed as part of the NIH Big Data To Knowledge (B2DK) initiative [1]. Subsequent work in the area cited the need for unique identifiers, supported by journals, and backed by extensive metadata [103].

## Examples

The landscape of metadata standards in tools is best organized into efforts to describe tools, dependencies, and containers.

### CRAN, EDAM, & CodeMeta - Tool description and citation

Source code spans both tools and literate statistical reports, although for convenience we classify code as a subcategory of tools. Metadata standards do not exist for loose code, but packaging manifests with excellent metadata standards exist for several languages, such as R's Comprehensive R Archive Network (CRAN) DESCRIPTION files (Figure 7).

```
Package: DESeq2
Type: Package
Title: Differential gene expression analysis based on the negative
    binomial distribution
Version: 1.27.31
Authors@R: c(
    person("Michael", "Love", email="michaelisaiahlove@gmail.com", role = c("aut","cre")),
    person("Constantin", "Ahlmann-Eltze", role = c("ctb")),
    person("Simon", "Anders", role = c("aut","ctb")),
    person("Wolfgang", "Huber", role = c("aut","ctb")))
Maintainer: Michael Love <michaelisaiahlove@gmail.com>
Description: Estimate variance-mean dependence in count data from
    high-throughput sequencing assays and test for differential
    expression based on a model using the negative binomial
    distribution.
License: LGPL (>= 3)
VignetteBuilder:
    knitr,
    rmarkdown
Imports: BiocGenerics (>= 0.7.5), Biobase, BiocParallel, genefilter,
    methods, stats4, locfit, geneplotter, ggplot2, Rcpp (>= 0.11.0)
Depends: S4Vectors (>= 0.23.18), IRanges, GenomicRanges,
    SummarizedExperiment (>= 1.1.6)
Suggests: testthat, knitr, rmarkdown, vsn, pheatmap, RColorBrewer,
    apeglm, ashr, tximport, tximeta, tximportData, readr, pbapply,
    airway, pasilla (>= 0.2.10)
LinkingTo: Rcpp, RcppArmadillo
URL: https://github.com/mikelove/DESeq2
biocViews: Sequencing, RNASeq, ChIPSeq, GeneExpression, Transcription,
    Normalization, DifferentialExpression, Bayesian, Regression,
    PrincipalComponent, Clustering, ImmunoOncology
RoxygenNote: 6.1.1
Encoding: UTF-8
```

*Figure 7. An R package DESCRIPTION file from DESeq2* [104]

Recent developments in tools metadata have focused on tool description, citation, dependency management, and containerization. The last two advances, exemplified by the Conda and Docker projects (described below), have largely made computational reproducibility possible, at least in the narrow sense of being able to reliably version and install software and related dependencies on other people's machines. Often small changes in software and reference data can have significant effects on an analysis [105]. Tools like Docker and Conda respectively make the computing environment and version pinning software tenable, thereby producing portable and stable environments for reproducible computational research.

The EMBRACE Data And Methods (EDAM) ontology provides high-level descriptions of tools, processes, and biological file formats [60]. It has been used extensively in tool recommenders [106], tool registries [107], and within pipeline frameworks and workflow languages [108,109]. In the context of workflows, certain tool combinations tend to be chained in predictable usage patterns driven by application; these patterns can be mined for tool recommender software used in workbenches [110].

CodeMeta [111] prescribes JSON-LD (JSON for Linked Data) standards for code metadata markup. While CodeMeta is not itself an ontology, it leverages Schema.org ontologies to provide language-agnostic means of describing software as well as "crosswalks" to translate manifests from various software repositories, registries, and archives into CodeMeta (Figure 8).

```
{
  "@context": [
    "https://doi.org/10.5063/schema/codemeta-2.0",
    "http://schema.org"
  ],
  "@type": "SoftwareSourceCode",
  "identifier": "baydem",
  "description": "Bayesian tools for reconstructing past and present\n    demography. Th
  "name": "baydem: Bayesian Tools for Reconstructing Past and Present\n    Demography",
  "license": "https://spdx.org/licenses/MIT",
  "version": "0.1.0",
  "programmingLanguage": {
    "@type": "ComputerLanguage",
    "name": "R",
    "version": "3.6.3",
    "url": "https://r-project.org"
  },
  "runtimePlatform": "R version 3.6.3 (2020-02-29)",
```

*Figure 8: A snippet of CodeMeta JSON file from Price et al. [112] using Schema.org contextual tags*

Considerable strides have been made in improving software citation standards [113], which should improve the provenance of methods sections that cite those tools that do not already have accompanying manuscripts. Code attribution is implicitly fostered by the application of large-scale data mining of code repositories such as Github is the generation of dependency networks [114], measures of impact [115], and reproducibility censuses [116].

### Dependency and package management metadata

Compiled software often depends on libraries that are shared by many programs on an operating system. Conflicts between versions of these libraries, and software that demands obscure or outdated versions of these libraries, are a common source of frustration for users who install scientific software and a major hurdle to distributing reproducible code. Until recently, installation woes and "dependency hell" were considered a primary stumbling block to reproducible research [117]. Software written in high-level languages such as Python and R has traditionally relied on language-specific package management systems and repositories, e.g., pip and PyPI for Python, and the install.packages() function and CRAN for R. The complexity yet unavoidability of controlling dependencies led to competing and evolving tools, such as pip, Pipenv, and Poetry in the Python community, and even different conceptual approaches, such as the CRAN time machine. In recent years, a growing number of scientific software projects utilize combinations of Python and compiled software. The Conda project (https://conda.io) was developed to provide a universal solution for compiled executables and script dependencies written in any language. The elegance of providing a single requirements file has contributed to Conda's rapid adoption for domain-specific library collections such as Bioconda [118], which are maintained in "channels" which can be subscribed and prioritized by users.

### Fledgling standards for containers

For software that requires a particular environment and dependencies that may conflict with an existing setup, a lightweight containerization layer provides a means of isolating processes from the underlying operating system, basically providing each program with its own miniature operating system. The ENCODE project [119] provided a virtual machine for a reproducible analysis that produced many figures featured in the article and serves as one of the earliest examples of an embedded virtual environment. While originally designed for deploying and testing e-commerce web applications, the Docker containerization system has become useful for scientific environments where dependencies and permissions become unruly. Several papers have demonstrated the usefulness of Docker for reproducible workflows [117] [120] and as a central unit of tool distribution [121,122].

Conda programs can be trivially Dockerized, and every BioConda package gets a corresponding BioContainer [123] image built for Docker and Singularity, a similar container solution designed for research environments. Because Dockerfiles are similar to shell scripts, Docker metadata is an underutilized resource and one that may need to be further leveraged for reproducibility. Docker does allow for arbitrary custom key-value metadata (labels) to be embedded in containers (Figure 9). The Open Container Initiative's Image Format Specification (https://github.com/opencontainers/image-spec/) defines pre-defined keys, e.g., for authorship, links, and licenses. In practice, the now deprecated Label Schema (http://label-schema.org/rc1/) labels are still pervasive, and users may add arbitrary labels with prepended namespaces. It should be noted that containerization is not a panacea and Dockerfiles can introduce irreproducibility and decay if contained software is not sufficiently pinned (e.g., by using so-called lockfiles) and installed from sources that are available in the future.

```
LABEL maintainer="daniel.nuest@uni-muenster.de" \
  Name="Reproducible research at GIScience - computing environment" \
  org.opencontainers.image.created="2020-04" \
  org.opencontainers.image.authors="Daniel Nüst" \
  org.opencontainers.image.url="https://github.com/nuest/reproducible-research-at-giscience/blob/master/Dockerfile" \
  org.opencontainers.image.documentation="https://github.com/nuest/reproducible-research-at-giscience/" \
  org.opencontainers.image.licenses="Apache-2.0" \
  org.label-schema.description="Reproducible workflow image (license: Apache 2.0)"
```

*Figure 9: Excerpt from a Dockerfile: LABEL instruction with image metadata, source: [https://github.com/nuest/ten-simple-rules-dockerfiles/blob/master/examples/text-analysis-wordclouds_R-Binder/Dockerfile](https://github.com/nuest/ten-simple-rules-dockerfiles/blob/master/examples/text-analysis-wordclouds_R-Binder/Dockerfile)*

## Future directions

### Automated repository metadata

Source code repositories such as Github and Bitbucket are designed for collaborative development, version control, and distribution and as such do not enforce any reproducible research standards that would be useful for evaluating scientific code submissions. As a corresponding example to the NLP above, there are now efforts to mine source code repositories for discovery and reuse [124].

### Data as a dependency

*Data libraries*, which pair data sources with common programmatic methods for querying them are very popular in centralized open source repositories such as Bioconductor [125], and scikit-learn [126], despite often being large downloads. Tierney and Ram provide a best practices guide to the organization and necessary metadata for data libraries and independent data sets [127]. Ideally, users and data providers should be able to distribute data recipes in a decentralized fashion, for instance, by broadcasting data libraries in user channels. Most raw data includes a limited number of formats, but ideally, data should be distributed in packages bound to a variety of tested formatters. One solution, Gogetdata [128] is a project that can be used to specify versioned *data* prerequisites to coexist with software within the Conda requirements specification file. A private company called Quilt is developing similar data-as-a-dependency solutions bound to a cloud computing model. A similar effort, Frictionless Data, focuses on JSON-encoded schemas for tabular data and data packages featuring a manifest to describe constitutive elements. From a Docker-centric perspective, the Open Container Initiative [129] is working to standardize "filesystem bundles" - the collection of files in a container and their metadata. In particular, container metadata is critical for relating the contents of a container to its source code and version, its relationship with other containers, and how to use the container.

Neither Conda nor Docker is explicitly designed to describe software with fixed metadata standards or controlled vocabularies. This suggests that a centralized database should serve as a primary metadata repository for tool information - rather than a source code repository, package manager, or container store. An example of such a database is the GA4GH Dockstore [130], a hub and associated website that allows for a standardized means of describing and invoking Dockerized tools as well as sharing workflows based on them.

## 3. Statistical reports & Notebooks

Statistical reports and notebooks serve as an annotated session of an analysis. Though they typically use input data that has been processed by scripts and workflows (layer 4 below), they can be characterized as a step in the workflow rather than apart from it, and for some smaller analyses, all processing can be done within these notebooks. Statistical reports and notebooks occupy an elevated reputation as being an exemplar of reproducible best practices, but they are not a reproducibility panacea and can introduce additional challenges - one reason being the metadata supporting them is surprisingly sparse.

Statistical reports which utilize *literate programming*, combining statistical code with descriptive text, markup, and visualizations have been a standard for statistical communication since the advent of Sweave [131]. Sweave allowed R and LaTeX markup to be mixed in chunks, allowing the adjacent contextual descriptions of statistical code to serve as guideposts for anyone reading a Sweave report, typically rendered as PDF. An evolution of Sweave, knitr [132], extended choices of both markup (allowing Markdown) and output (HTML) while enabling tighter integration with integrated development environments such as RStudio [133]. A related project which started in the Python ecosystem but now supports several kernels, Jupyter [134], combined the concept of literate programming with a REPL (read-eval-print loop) in a web-based interactive session in which each block of code is kept stateful and can be re-evaluated. These live documents are known as "notebooks." Notebooks provide a means of allowing users to directly analyze data programmatically using common scripting languages, and access more advanced data science environments such as Spark, without requiring data downloads or localized tool installation if run on cloud infrastructures. Using preloaded libraries, cloud-based notebooks can alleviate time-consuming permissions recertification, downloading of data, and dependency resolution, while still allowing persistent analysis sessions. Data-set specific Jupyter notebooks "spawned" for thousands of individuals temporarily have been enabled as companions for Nature articles [135] and are commonly used in education. Cloud-based notebooks have not yet been extensively used in data portals, but they represent the analytical keystone to the decade-long goal of "bringing the tools to the data." Notebooks offer possibilities over siloed installations in terms of eliminating the data science bottlenecks common to data analyses - cloud-based analytic stacks, cookbooks, and shared notebooks.

Collaborative notebook sharing has been used to accelerate the analysis cycle by allowing users to leverage existing code. The predictive analytics platform Kaggle employs an open implementation of this strategy to host data exploration events. This approach is especially useful for sharing data cleaning tasks - removing missing values, miscategorizations, and phenotypic standardization which can represent 80% of effort in an analysis [136]. Sharing capabilities in existing open-source notebook platforms are at a nascent stage, but this presents significant possibilities for reproducible research environments to flourish. One promising project in this area is Binder, which allows users to instantiate live Jupyter notebooks and associated Dockerfiles stored on Github within a Kubernetes-backed service [137,138].

At face value, reports and notebooks resemble source code or scripts, but as the vast majority of statistical analysis and machine learning education and research is conducted in notebooks, they represent an important area for reproducibility.

## Examples

### RMarkdown headers

As we mentioned, statistical reports and notebooks do not typically leverage structured metadata for reproducibility. R Markdown based reports, such as those processed by knitr, do have a YAML-based header (Figure 10). These are used for a wide variety of technical parameters for controlling display options, for providing metadata on authors, e.g., when used for scientific publications with the *rticles* package [139], or for parameterizing the included workflow (https://rmarkdown.rstudio.com/developer_parameterized_reports.html%23parameter_types%2F). However, no schema or standards exist for their validation.

```
---
title: "A title for the analysis"
output:
  html_document:
    theme: lumen
    toc: true
    toc_float:
      collapsed: false
    code_folding: show
---
```

*Figure 10: A YAML-based RMarkdown header from https://github.com/jmonlong/MonBUG18_RMarkdown*

### Statistical and Machine Learning Metadata Standards

The intense interest paired with the competitive nature of machine learning and deep learning conferences such as Neurips demands high reproducibility standards [140]. Given the predominance of notebooks for disseminating machine learning workflow, we focused our attention on finding statistical and machine learning metadata standards that would apply to content found with notebooks. The opacity, rapid proliferation, and multifaceted nature of machine learning and data mining statistical methods to non-experts suggest it is necessary to begin cataloguing and describing them at a more refined level than crude categories (e.g. clustering, classification, regression, dimension reduction, feature selection). So far, the closest attempt to decompose statistics in this manner is the STATO statistical ontology

(http://stato-ontology.org/), which can be used to semantically, rather than programmatically or mathematically, define all aspects of a statistical model and its results, including assumptions, variables, covariates, and parameters (Figure 11). While STATO is currently focused on univariate statistics, it represents one possible conception for enabling broader reproducibility than simply relying on specific programmatic implementations of statistical routines.

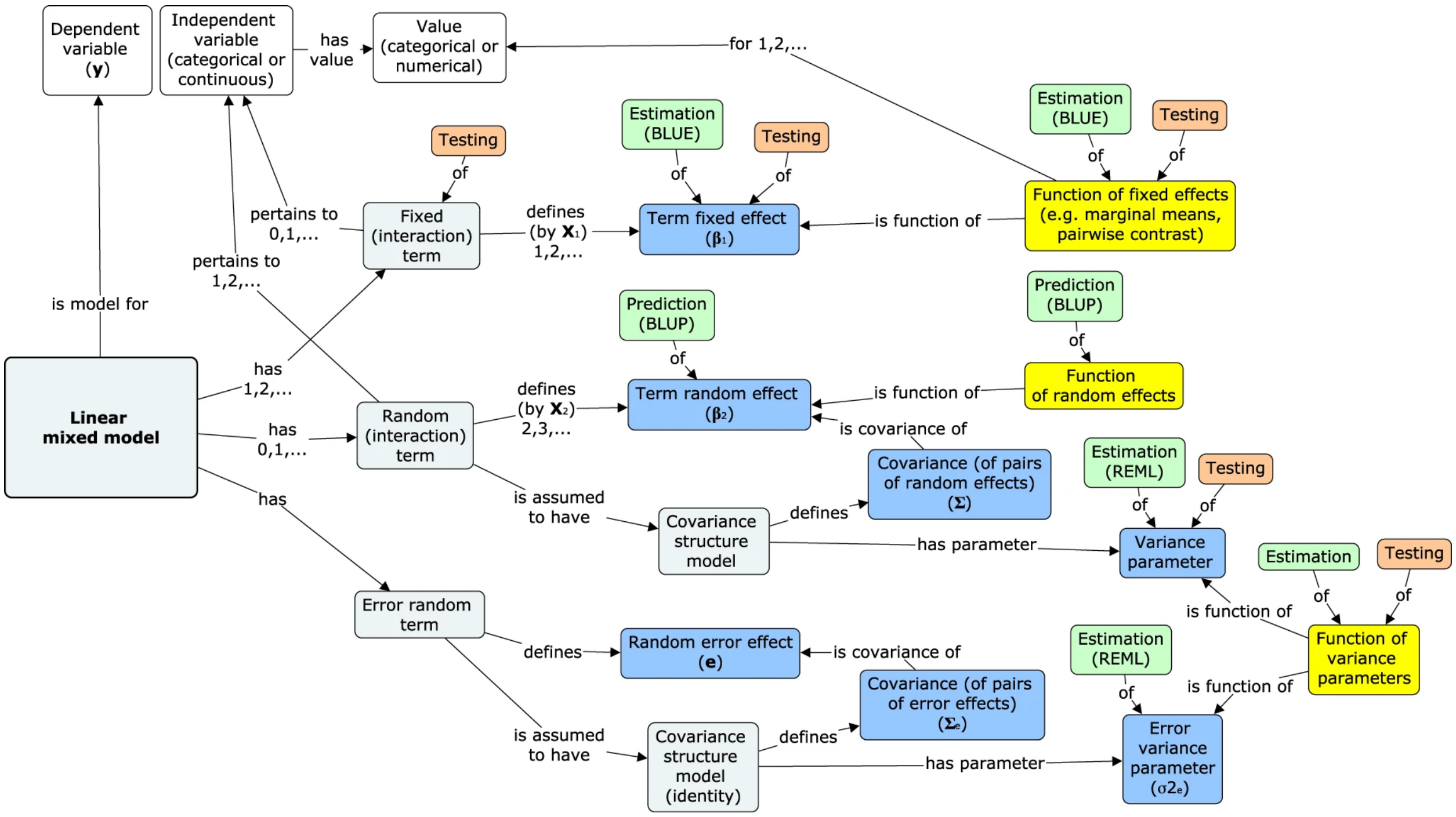

*Figure 11: Concepts describing a linear mixed model used by STATO* [141]

MEX is designed as a vocabulary to describe the components of machine learning workflows. The MEX vocabulary builds on PROV-O to describe specific machine learning concepts such as hyperparameters and performance measures and includes a decorator class to work with Python.

## Future directions - parameter tracking

MLFlow [142] is designed specifically to handle hyperparameter tracking for machine learning iterations or "runs" performed in the Apache Spark , but also tracks arbitrary artifacts and metrics associated with these. The metadata format MLFlow uses exposes variables which are explored and tuned by end-users (Figure 12).

```
name: HyperparameterSearch

conda_env: conda.yaml

entry_points:
  # train Keras DL model
  train:
    parameters:
      training_data: {type: string, default: "../sklearn_elasticnet_wine/wine-quality.csv"}
      epochs: {type: int, default: 32}
      batch_size: {type: int, default: 16}
      learning_rate: {type: float, default: 1e-1}
      momentum: {type: float, default: .0}
      seed: {type: int, default: 97531}
    command: "python train.py {training_data}
                                    --batch-size {batch_size}
                                    --epochs {epochs}
                                    --learning-rate {learning_rate}
                                    --momentum {momentum}"
```

*Figure 12: MLflow snippet showing exposed hyperparameters*

## 4. Pipelines

Most scientific analyses are conducted in the form of pipelines, in which a series of transformations is performed on raw data, followed by statistical tests and report generation. Pipelines are also referred to as "workflows," which sometimes also encompasses steps outside an automated computational process. Pipelines represent the computation component of many papers, in both basic research and tool papers. Pipeline frameworks or scientific workflow management systems (SWfMS) are platforms that enable the creation and deployment of reproducible pipelines in a variety of computational settings including cluster and cloud parallelization. The use of pipeline frameworks, as opposed to standalone scripts, has recently gained traction, largely due to the same factors (big data, big science) driving the interest of reproducible research. Although frameworks are not inherently more reproducible than shell scripts or other scripted ad hoc solutions, use of them tends to encourage parameterization and configuration that promote reproducibility and metadata. Pipeline frameworks are also attractive to scientific workflows in that they provide tools for the reentrancy - restarting a workflow where it left off, implicit dependency resolution - allowing the framework engine to automatically chain together a series of transformation tasks, or "rules," to produce a give a user-supplied file target. Collecting and analyzing provenance, which refers to the record of all activities that go into producing a data object, is a key challenge for the design of pipelines and pipeline frameworks.

The number and variety of pipeline frameworks have increased dramatically in recent years - each framework built with design philosophies that offer varying levels of convenience, user-friendliness, and performance. There are also tradeoffs between the dynamicity of a framework, in terms of its ability to behave flexibly (e.g. skip certain tasks, re-use results from a cache) based on input, that will affect the apparent reproducibility and the run-level metadata that is required to inspire confidence in an analyst's ability to infer how a pipeline behaved in a particular situation. Leipzig [143] reviewed and categorized these frameworks into three key dimensions: using an implicit or explicit syntax, using a configuration, convention or class-based design paradigm and offering a command line or workbench interface.

***Convention-based frameworks*** are typically implemented in a domain-specific language, a meaningful symbol set to represent rule input, output, and parameters that augment existing scripting languages to provide the glue to create workflows. These can often mix shell-executable commands with internal script logic in a flexible manner. **Class-based pipeline frameworks** augment programming languages to offer fine-granularity means of efficient distribution of data for high-performance cluster computing frameworks such as Apache Spark. ***Configuration based frameworks*** abstract pipelines into configuration files, typically XML or JSON which contain little or no code. Workbenches such as Galaxy [144], Kepler [145], KNIME [146], Taverna [147], and commercial workbenches such as Seven Bridges Genomics and DNANexus typically offer canvas-like graphical user interfaces by which tasks can be connected and always rely on configuration-based tool and workflow descriptors. Customized workbenches configured with a selection of pre-loaded tools and workflows and paired with a community web portal are often termed "science gateways."

## Examples

### CWL - A configuration-based framework for interoperability

The Common Workflow Language (CWL) [148] is a specification for tools and workflows to share across several pipeline frameworks, adopted by several workbenches. CWL manages the exacting specification of file inputs, outputs, parameters that are "operational metadata" - used by the workflow machinery to communicate with the shell and executable software (Figure 13). While this metadata is primarily operational in nature and rarely accessed outside the context of a compatible runner such as Rabix [149] or Toil [150], CWL also enables a tool metadata in the form of versioning, citation, and vendor-specific fields that may differ between implementations.

Using this metadata, an important aspect of CWL is the focus on richly describing tool invocations both for reproducibility and documentation purposes, with tools referenced as retrievable Docker images or Conda packages, and identifiers to EDAM [60], ELIXIR's bio.tools [56] registry and Research Resource Identifiers (RRIDs) [151]. This wrapping of command line tool interfaces is used by GA4GH Dockstore [130] for providing a uniform executable interface to a large variety of computational tools even outside workflows.

While there are many other configuration-based workflow languages, CWL is notable for the number of parsers that support its creation and interpretation, and an advanced linked data validation language, called Schema Salad. Together with supporting projects, such as Research Objects, the CWL appears amenable to being used as metadata.

```
#!/usr/bin/env cwl-runner
cwlVersion: v1.0
class: Workflow

requirements:
  StepInputExpressionRequirement: {}

doc: |
  Author: AMBARISH KUMAR er.ambarish@gmail.com & ambari73_sit@jnu.ac.in
  This is a proposed standard operating procedure for genomic variant detection using GATK4.
  It is hoped to be effective and useful for getting SARS-CoV-2 genome variants.

  It uses Illumina RNASEQ reads and genome sequence.

inputs:
  sars_cov_2_reference_genome:
    type: File
    format: edam:format_1929  # FASTA

  rnaseq_left_reads:
    type: File
    format: edam:format_1930  # FASTQ

  rnaseq_right_reads:
    type: File
    format: edam:format_1930  # FASTQ

steps:
  index_reference_genome_with_bowtie2:
    run: ../tools/bowtie2/bowtie2_build.cwl
    in:
      reference_in: sars_cov_2_reference_genome
      bt2_index_base:
        valueFrom: "sars-cov-2"
    out: [ indices ]
```

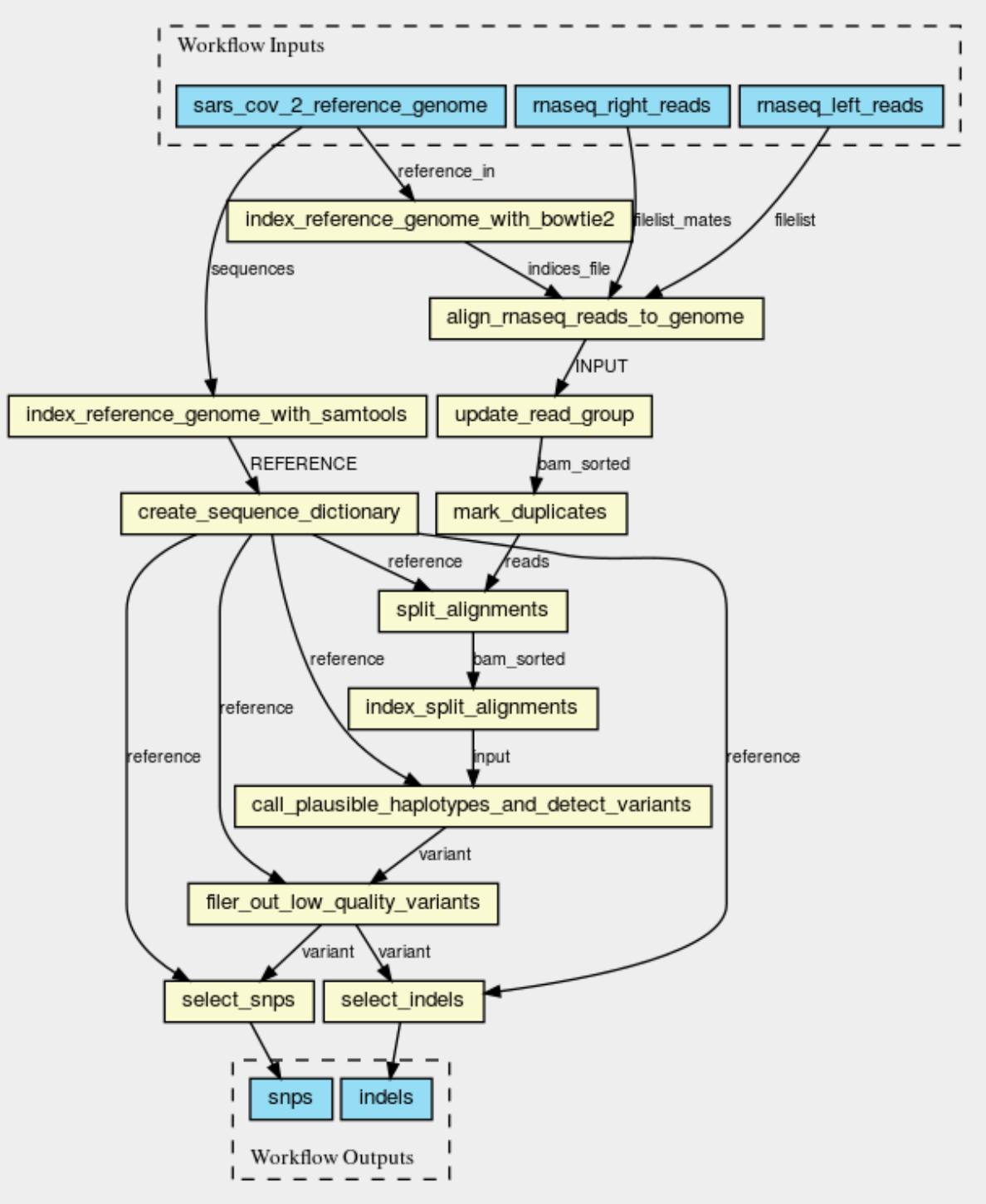

*Figure 13: Snippets of a COVID-19 variant detection CWL workflow and the workflow as viewed through the cwl-viewer [152]. Note the EDAM file definitions.*

## Future directions

### Interoperable script and workflow provenance

For future metadata to support pipeline reproducibility must accommodate huge menagerie of solutions that coexist inside a number of computing environments. Large organizations have been encouraging the use of cloud-based data commons, but solutions that target the majority of published scientific analysis must address the fact that many if not most of them will not use a data commons or even a pipeline framework. Because truly reproducible research implies evaluation by third parties, portability is an ongoing concern.

Pimental et al. reviewed and categorized 27 approaches to collecting provenance from scripts [153]. A wide variety of relational databases and proprietary file formats are used to store, distribute, visualize, version, and query provenance from these tools. The authors found while four approaches - RDataTracker [154], SPADE [155], StarFlow [156], and YesWorkflow [157] - natively adopt interoperable W3C PROV or OPM standards as export, most were designed for internal usage and did not enable sharing or comparisons of provenance. In part, these limitations are related to primary goals and scope of these provenance tracking tools.

For analyses that use workflows, a prerequisite for reproducible research is the ability to reliably share "workflow enactments," or runs which encompass all elements of the analytic stack. Unlike pipeline frameworks geared toward cloud-enabled scalability, compatibility with executable command-line arguments and programmatic extensibility afforded by DSLs, Vistrails was designed explicitly to foster provenance tracking and querying, both prospective and retrospective [158]. As part of the WINGS project, Garijo [159] uses linked-data standards - OWL, PROV, and RDF to create a framework-agnostic Open Provenance for Workflows (OPMW) for greater semantic possibilities for user needs in workflow discovery and publishing. The CWLProv [72] project implements a CWL-centric and RO-based solution with a goal of defining a format of implementing retrospective provenance.

### Packaging and binding building blocks

While we have attempted to classify metadata across layers of the analytic stack, there are a number of efforts to tie or bind all these metadata that define a research compendia explicitly. A Research Compendium (RC) is a container for building blocks of a scientific workflow. Originally defined by Gentleman and Temple Lang as a means for distributing and managing documents, data, and computations using a programming language’s packaging mechanism, the term is now used in different communities to provide code, data, and documentation (including scientific manuscripts) in a meaningful and useable way (https://research-compendium.science/). A best practice compendium includes environment configuration files (see above), has files are under version control and uses accessible plain text formats. Instead of a formal workflow specification, inputs, outputs, and control files and the required commands are documented for human users in a README file. While an RC can take many forms, the flexibility is also a challenge for extracting metadata. The Executable Research Compendium (ERC) formalizes the RC concept with an R Markdown notebook for the workflow and a Docker container for the runtime environment [160]. A YAML configuration file connects these parts, configures the document to be displayed to a human user, and provides minimal metadata on licenses. The concept of bindings connects interactive parts of and ERC workflow with the underlying code and data[161].

```yaml
id: b9b0099e-9f8d-4a33-8acf-cb0c062efaec
spec_version: 1
main: workflow.Rmd
display: paper.html
licenses:
    code: Apache-2.0
    data: data-licenses.txt
    text: "Creative Commons Attribution 2.0 Generic (CC BY 2.0)"
    metadata: "see metadata license headers"
```

*Figure 14: erc.yml example file, see the specification at https://o2r.info/erc-spec/.*

Instead of trying to establish a common standard and single point for metadata, the ERC intentionally skips formal metadata and exports the known information into multiple output files and formats, such as Zenodo metadata as JSON or Datacite as XML, accepting duplication for the chance to provide usable information in the long term.

Perhaps the most prominent realization of the RC concept are Research Objects [162] and the subsequent RO-Crate [163] projects, which strive to be comprehensive solutions for binding code, data, workflows, and publications into a metadata-defined package. RO-Crate is lightweight JSON-LD (javascript object notation linked data) which supports Schema.org concepts to identify and describe all constituent files from the analytic stack and various people, publication, and licensing metadata, as well as provenance both between workflows and files and across crate versions.

```
{ "@context": "https://w3id.org/ro/crate/1.0/context",  
  "@graph": [

 {
    "@type": "CreativeWork",
    "@id": "ro-crate-metadata.jsonld",
    "conformsTo": {"@id": "https://w3id.org/ro/crate/1.0"},
    "about": {"@id": "./"}
 },  
 {
    "@id": "./",
    "identifier": "https://doi.org/10.4225/59/59672c09f4a4b",
    "@type": "Dataset",
    "datePublished": "2017",
    "name": "Data files associated with the manuscript:Effects of facilitated family case conferencing for ...",
    "description": "Palliative care planning for nursing home residents with advanced dementia ...",
    "license": {"@id": "https://creativecommons.org/licenses/by-nc-sa/3.0/au/"}
 },
{
  "@id": "https://creativecommons.org/licenses/by-nc-sa/3.0/au/",
  "@type": "CreativeWork",
  "description": "This work is licensed under the Creative Commons Attribution-NonCommercial-ShareAlike ...",
  "identifier": "https://creativecommons.org/licenses/by-nc-sa/3.0/au/",
  "name": "Attribution-NonCommercial-ShareAlike 3.0 Australia (CC BY-NC-SA 3.0 AU)"
  }
 ]
}
```

*Figure 15: RO-Crate metadata*

An alternative approach to binding is to leverage existing work in "application profiles" [164], a highly customizable means of combining namespaces from different metadata schemas. Application profiles follow along the Singapore Framework, and guidelines supported by the Dublin Core Metadata Initiative (DCMI).

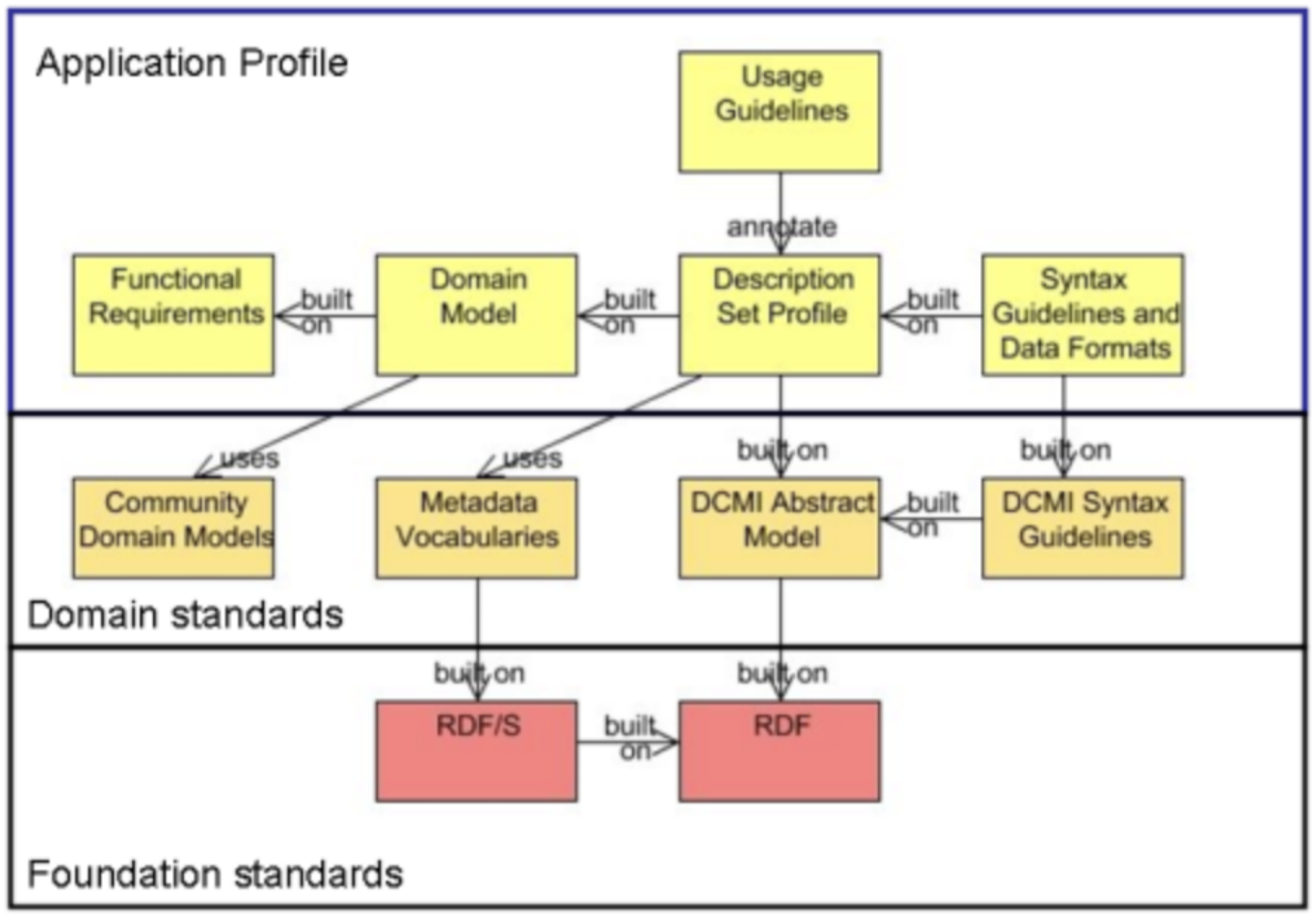

*Figure 16: Singapore Framework application profile model*

## 5. Publication

Our conception of the analytic stack points to the manuscript as the final product of an analysis. Due to the requirements of cataloging, publishing, attribution, and bibliographic management, journals employ a robust set of standards including MARC21 and ISO_2709 for citations, and Journal Article Tag Suite (JATS) for manuscripts. Library science has been an early adopter of many metadata standards and encoding formats (e.g. XML) later used throughout the analytic stack. Supplementing and extending these standards to accommodate reproducible analyses connected or even embedded in publications is an open area for development.

For the purposes of reproducibility we are most interested in finding publication metadata standards that attempt to support structured results as a "first-class citizen" - essentially input metadata but for integration into the manuscript.

The methods section of a peer-reviewed article is the oldest and often the sole source of metadata related to an analysis. However, methods sections and other free-text supplementals are notoriously poor and unreliable examples of reproducible computational research, as evidenced by the Amgen findings. A number of text mining efforts have sought to extract details of the software used in analyses directly from methods sections for purposes of survey [165,166] and recommendation [167] using natural language processing (NLP). The ProvCaRe database and web application extend this to both computational and clinical findings by using a wide-ranging corpus of provenance terms and extending existing PROV-O ontology [168]. While these efforts are noble, they can never entirely bridge the gap between human-readable protocols and RCR.

Journals share an important responsibility to enforce and incentivize reproducible research, but most peer-reviewed publications have been derelict in this role. While many have raised standards for open data access, "open analysis" is still an alien concept to many journals. Some journals, such as Nature Methods, do require authors to submit source code [169]. Of the most prestigious life science journals (Nature, Science, Cell), the requirements vary considerably and it is not clear how these guidelines are actually enforced [170]

Container portals, package repositories, and workbenches do provide some additional inherent structure that would be useful for journals to require, but these often lack any binding with notebooks or elegant routes to report generation that would guarantee the scientific code matches the results contained with a manuscript. Computational provenance between all figures and tables in a manuscript and the underlying analysis is an open area of research that we discuss below.

### Examples

### Formalization of the Results of Biological Discovery

In the scientific literature, authors must not only outline the formulation of their experiments, their execution, and their results, but also an interpretation of the results with respect to an overarching scientific goal. Due to the lack of specificity of prose and the needless jargon endemic to modern scientific discourse, both the goals and interpretation of results are often obfuscated such that the reader must exert considerable effort to understand. This burden is further exacerbated by the acceleration of the growth of the body of scientific literature. As a result, it has become overwhelming, if not impossible, for researchers to follow the relevant literature in their respective fields, even with the assistance of search tools like PubMed and Google.

The solution lies in the formalization of the interpretation presented in the scientific literature. In molecular biology, several formalisms (e.g. BEL [85], SBML [171], SBGN [172], BioPAX [173], GO-CAM [174]) have the facility to describe the interactions between biological entities that are often elucidated through laboratory or clinical experimentation. Further, there are several organizations [175–179] whose purpose is to curate and formalize the scientific literature in these formats and distribute them in one of several databases and repositories. Because curation is both difficult and time-consuming, several semi-automated NLP [180,181] curation workflows based on NLP-based relation extraction systems [182–184] and assemblers [185] have been proposed to assist.

The Biological Expression Language (BEL) captures causal, correlative, and associative relationships between biological entities along with the experimental/biological context in which they were observed as well as the provenance of the publication from which the relation was reported (https://biological-expression-languge.github.io). It uses a text-based custom domain-specific language (DSL) to enable biologists and curators alike to express the interpretations present in biomedical texts in a simple but structured form, as opposed to a complicated formalism built with low-level formats XML, JSON, and RDF or mid-level formats like OWL and OBO. Similarly to OWL and OBO, BEL pays deep respect to the need for the use of structured identifiers and controlled vocabularies for its statements to support the integration of multiple content sources in downstream applications. We focus on BEL because of its unique ability to represent findings across biological scales, including the genomic, transcriptomic, proteomic, pathway, phenotype, and organism levels.

Below is a representation of a portion of the MAPK signaling pathway in BEL, which describes the process through which a series of kinases are phosphorylated, become active, and phosphorylate the next kinase in the pathway. It uses the FamPlex (fplx*)* [186]namespace to describe the RAF, MEK, and ERK protein families.

```
act(p(fplx:RAF), ma(kin)) directlyIncreases     p(fplx:MEK, pmod(Ph))
    p(fplx:MEK, pmod(Ph)) directlyIncreases act(p(fplx:MEK), ma(kin))
act(p(fplx:MEK), ma(kin)) directlyIncreases     p(fplx:ERK, pmod(Ph))
```

```
p(fplx:ERK, pmod(Ph)) directlyIncreases act(p(fplx:ERK)))
```

*Figure 17: MAPK signalling pathway in Biological Expression Language (BEL)*

While the additional provenance, context, and metadata associated with each statement have not been shown, this example demonstrates that several disparate information sources can be assembled in a graph-like structure due to the triple-like nature of BEL statements.

While BEL was designed to express the interpretation presented in the literature, related formats are more focused on mechanistically describing the underlying processes on either a qualitative (e.g., BioPAX, SBGN) or quantitative (e.g., SBML) basis. Ultimately, each of these formalisms has supported a new generation of analytical techniques that have begun to replace classical pathway-analysis.

### Future directions - reproducible articles

Attempts have been made to integrate reproducible analyses into manuscripts. An article in eLife [187] was published with an inline live RMarkdown Binder analysis as part of a proof-of-concept of the publisher's Reproducible Document Stack (RDS) [188]. Because of the technical metadata used for rendering and display, subtle changes are required to integrate containerized analyses with JATS, and the requirements for hosting workflows outside the narrow context of Binder will require further engineering and metadata standards.

## Discussion

The range and diversity of metadata standards developed to aid researchers in their daily activities, but also in sharing research (data, code, publications), and contributing to open science is extensive. If we promote metadata as the "glue" of reproducible research, what does that entail for the metadata and RCR communities? While there are overlapping standards, and no one metadata schema can support all aspects of the analytic stack, it is important to recognize that the metadata developments pursued, particularly the standards that are shared and maintained by a community, many of which have gone through formal standards review processes, demonstrate value to their communities. Science has no boundary, and while these standards may have been developed to meet more specific needs as part of the research life-cycle, as reviewed above, they have a continuing value for RCR.

In our review, we have attempted to describe metadata as it addresses reproducibility across the analytic stack. Two principal components: 1)Embeddedness vs connectedness and the 2) methodology weight and standardization appear to be recurring themes across all RCR metadata facets.

## Embeddedness vs connectedness

Certain efforts in the metadata context lend to the stickiness of experimental details from data collection to publications, and others are more directed to the goals of data sharing and immediate access. Data formats have an influence on the long-term reproducibility of analyses and reusability of input data, though these goals are not always aligned. Some binary data formats lend them to easily accommodate embedded metadata - i.e. metadata that is bound to its respective data by residing in the same file. In the case of the DICOM format used in medical imaging, a well-vetted set of instrumentation metadata is complemented by support for application-specific metadata. Downstream this has enabled support for DICOM images in various repositories such as The Cancer Imaging Archive [189]. The continued increase in the use of such imaging data has led to efforts to further leverage biomedical ontologies in tags [190] and issue DOIs to individual images [191]. As discussed above, the lack of support for complex metadata structures has not significantly hindered the adoption of DICOM for a variety of uses not anticipated by its authors (DICOM introduced in 1985). This could be an argument that embeddedness is more important than complexity for long-term sustainability, or merely that early arrivals tend to stay entrenched. In the case of Binary Alignment Map [192] files used to store genomic alignments, file-level metadata resides in an optional comment section above data. Once again, these are arbitrary human-readable strings with no inherent advanced data structure capabilities. In some instances, instrumentation can aid in reproducibility by embedding crucial metadata (such as location, instrument identifiers, and various settings) in such embedded formats with no manual input, although ideally this should be not simply be used at face value as a sanity check against metadata used in the analysis, for instance, to identify potential sample swaps or other integrity issues. Reliance on ad-hoc formatting methods of supporting extensibility, as in through serializations using commas or semicolons delimiters, can have deleterious effects on the stability of a format. In bioinformatics, a number of genomic position-based tabular file formats have faced "last-column bloat," as new programs have piled on an increasingly diverse array of annotations.

This rigid embedded scheme employed by DICOM stands in contrast to standards such as EML, where contributors are encouraged with a flexible ontology to support supplemental metadata for the express purposes of data sharing. MIAME appears to lie somewhere in the middle, where there is a required minimal subset of tags to be supplied, much of it from the microarray instruments itself and aided by a strong open source community (Bioconductor), and paired with a data availability incentive in order to publish associated manuscripts.

In terms of reproducibility, embeddedness represents a double-edged sword. As a packaging mechanism, embedded metadata serves to preserve aspects of attribution, provenance, and semantics for the sharing of individual files but a steadfast reliance on files can lead to siloing which may be antithetical to discovery (the "Findable" in FAIR). Files as the sole means of research data distribution are also contrary to the recent proliferation of "microservices" - Software-as-a-Service often instantiated in a serverless architecture and offering APIs. While provenance can be embedded in the headers described above, these types of files are more likely to be found at the earlier stages of an analysis, suggesting there is work to be done in developing embedded metadata solutions for notebook and report output if this is to be a viable general scheme. So much of reproducibility depends on the relay of provenance *between* layers of the analytic stack that the implementation of metadata should be optimized to encourage usage by the tools explored in this review.

Metadata is, of course, critical to the functioning of services that support the "semantic web," in which data on the world wide web is given context to enable it to be directly queried and processed, or "machine-readable." Several technologies enabling the semantic web and linked data RDF, OWL, SKOS, SPARQL, and JSON-LD are best recognized as metadata formats themselves or languages for metadata introspection allowing the web to behave like a database rather than a document store. Semantic web services now exist for such diverse data sources as gene-disease interactions [193] and geospatial data [194]. RDF triples are the core of knowledge graph projects such as DBpedia [195] and Bio2RDF [196]. The interest in using knowledge graphs for modeling and prediction in various domains, and the increased use of "embedding knowledge graphs," graph to vector transformations designed to augment AI approaches [197], has exposed the need for reproducibility and metadata standards in this area [198].

The development of large multi-institutional data repositories that characterize "big science" and remote web services that support both remote data usage and the vision of "bringing the tools to the data" make the cloud an appealing replacement for local computing resources [199]. This dependence on data and services hosted by others, however, introduces the threat of "workflow decay" [200] that requires extensive provenance tracking to freeze inputs and tools in order to ensure reproducibility at a later date.

The promise of distributed annotation services, automated discovery, and the integration of disparate forms of data, using web services and thereby avoiding massive downloads, is of central import to many areas of research. However, the import of the semantic web to RCR is a two-sided coin. On one hand, as noted by Aranguren and Wilkinson [201], the semantic web provides a formalized means providing context to data, which is a crucial part of reproducibility. The semantic web is by its very nature, open, and provides a universal low barrier to data access with few dependencies other than an internet connection. Conversely, a review of the semantic web's growing impact on cheminformatics [202] notes that issues of data integrity and provenance are of concern when steps in an analysis rely on data fetched piecemeal via a web service.

They provide a common source reference point for several unrelated analyses, but that can serve as a critical point of failure should they disappear. Projects serving to provide long-term archival solutions for scientific analyses need to cache or download webservice data. Along the same lines, often studies are conducted entirely from dedicated databases - relational, so-called "NoSQL" solutions - key-value, document stores, or column-stores. These can introduce substantial issues to portability and reproducibility, especially when studies access relational joins across subsets of these databases.

## Methodology weight and standardization

Our review has spotlighted several metadata solutions across a spectrum of heavyweight vs lightweight solutions, bespoke vs standard solutions and offering different levels of granularity, and adoption. Because these choices can often largely reflect those of the stakeholders involved in the design and their goals rather than immediate needs, a discussion of those groups is warranted.

### Sphere of Influence

Governing and standards-setting organizations (e.g. NIH, GA4GH, W3C) new applications (e.g. machine learning, translational health) and trends in the greater scientific community (open science, reproducible research) are steering metadata for reproducible research in different and broader directions than traditional stakeholders, individual researchers. There are also differences in the approaches taken between different scientific fields, with the life sciences arguably more varied in both the size of projects and the level of standards than those physical sciences (e.g. LIGO). This does not discount the fact that much of the progress in metadata for RCR has been originally intended for purposes other than reproducibility, and often "bespoke," or custom-designed solutions to address the problem at hand for small labs or individual investigators. A good example is the tximeta Bioconductor package, which implements reference transcriptome provenance for RNA-Seq experiments, extending a number of popular transcript quantification tools with checksum-based tracking and identification [203]. While this is an elegant solution, tximeta is focused on one analysis pattern.

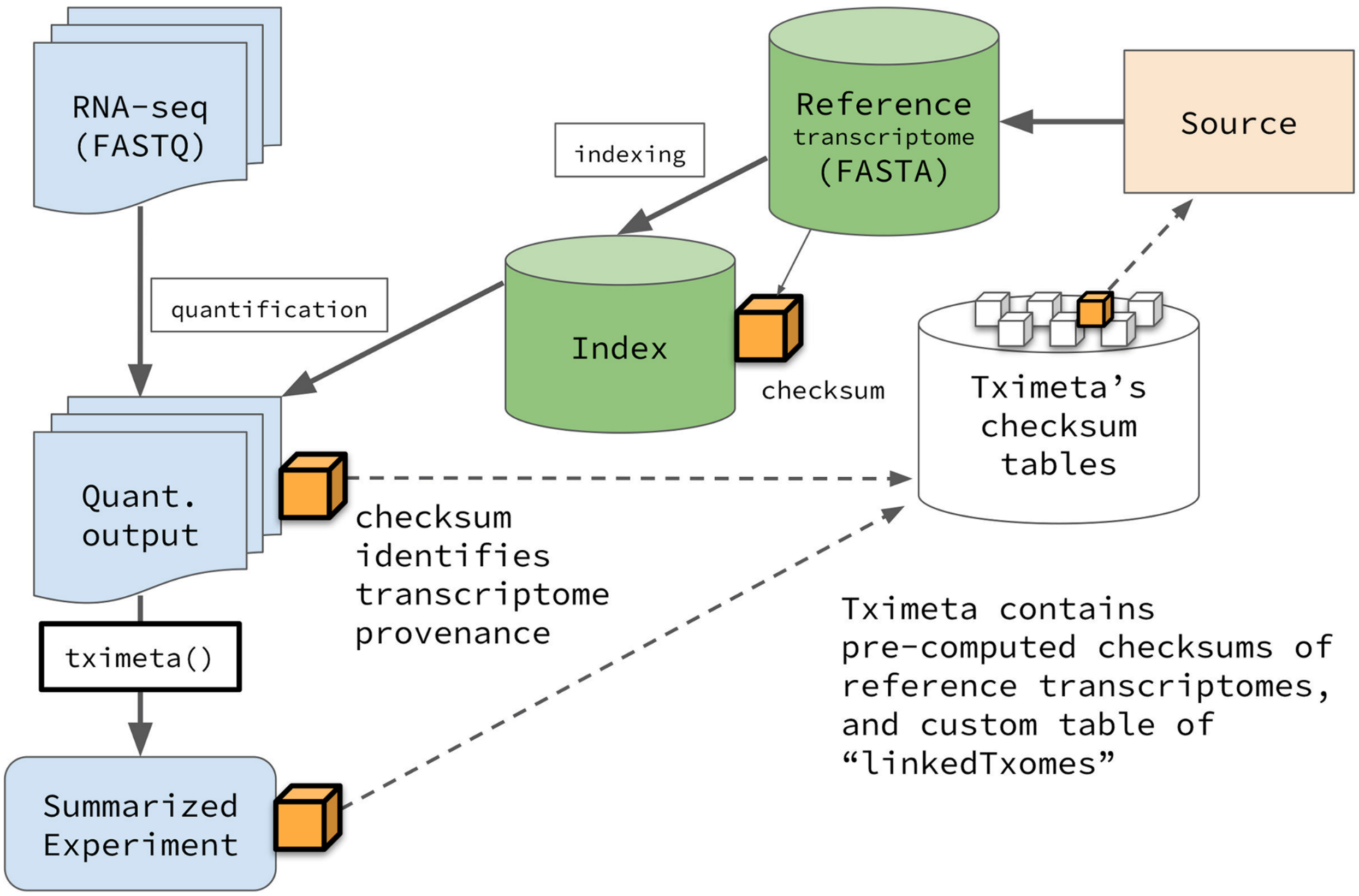

*Figure 18: The high-level schematic of tximeta* [203]

In practice, concerns over reproducibility appear to be correlated with the number of stakeholders. While there are highly conscientious scientists who have built tools and standards to support the reproducibility of their own work, pressure coming from attempts to reproduce published analyses, and the heightened reuse of data among participants in multi-institution consortia, data repositories, and data commons have forced the issue.

## Metadata capital and reuse

The term "metadata capital" [204] was coined to describe how an organization's efforts in producing high quality metadata can have a positive return on investment downstream. We contend this applies to RCR. In this context it may be useful to reposition the onus for collecting metadata along the competitiveness of smaller groups - labs, cores, and individual institutions. These smaller organizations clearly experience a reproducibility crisis in the form of impaired transfer of knowledge from outgoing to incoming trainees. However, the seminal Nature Baker survey of 1,500 scientists reported 34% of participants had not established procedures for reproducibility in their own labs [11]. Metadata reuse - for replication, generalization, meta-analyses, or general research use is enabled by RCR and elemental to FAIR. Reuse typically demands greater metadata needs than narrow-sense reproduction, for instance, to control for batch effects or various assumptions that go into original research [205]. Often the centralized submission portals demand more expansive metadata than an individual researcher would anticipate being necessary, belying their importance in the reproducibility and reuse process. Essential for reproducibility, surveys suggest provenance information is an important criteria for reuse in the physical sciences [206]. Metadata for data reuse has relevance for data harmonization for biomedical applications, such as toward highly granular phenotype ontologies, genotype-phenotype meta-analyses [207], generating synthetic controls for clinical trials, and, consent metadata such as the Data Use Ontology [208] to describe allowed downstream usage of patient data. Designing metadata for the needs of general reuse, especially outside narrow scientific domains, requires greater foresight than that needed for RCR but authors can follow similar templates.

## Recommendations & Future Work

Widespread adoption of RCR is highly dependent on a cultural shift within the scientific community [209,210] promoted by both journals and funding agencies. The allegorical "stick" of higher RCR standards should be accompanied by carrots in the form of publication incentives. One of these carrots could involve a support mechanism by which pre- and post-publication peer review can properly evaluate and test statistical methods cited in papers. Such a collaborative computational peer review could involve parameter exploration, swapping out individual statistical tests or tool components for similar substitutes, and using new data sets. An "advocated software peer review" enabled by RCR and conducted by reviewers taking a hands-on approach to strengthening analyses using collaborative interactive notebooks or other tools.[211]

One interesting development in this area is the growing interest in developing FAIR metrics and reproducibility "badges" to denote compliance. The FAIRshake toolkit implements rubrics to evaluate the digital resources such as datasets, tools, and workflows [212]. These rubrics include criteria such as data and code availability but also metadata such as contact information, description, and licensing embedded using Schema.org tags.

In terms of the analytic stack, there are several areas which offer low-hanging fruit for innovation. One is developing inline semantic metadata for publications and notebooks. While schema.org tags have been used for indexing data, to our knowledge there is no journal that supports, much less encourages, semantic markup of specific terms within a manuscript. There has been tacit support for such inline markup in newer manuscript composition tools such as Manubot [213], but generally Such terms could disambiguate concepts, point to the provenance of findings within a result section or from a figure, and accelerate linked data and discovery.

The lack of integration between notebooks and pipeline frameworks can create friction during the analysis process, which can discourage users from using them jointly. Efforts such as NoWorkflow/YesWorkflow and internal frameworks such as Targets [214] are helping to bridge these distinctions, but few solutions have sought to aid notebook-pipeline integration in general.

Secondly there is a clear need for greater annotation within statistical reports and notebooks for semantic markup to categorize and disambiguate machine learning and deep learning workflows. Because of the explosion in advances from this area, researchers outside the machine learning core community have found it difficult to keep up with the litany of terminology, techniques, and metrics being developed. Clearly metadata can play a role in augmenting users understanding of, for instance, existing technique relates most closely with a new one. This will ensure the broader goals of reproducibility.

Statistical metadata is vital for users to discover, and reviewers to evaluate complex statistical analyses [215], but metadata that describe statistical methods is largely non-existent. The increasing diversity and application of machine learning approaches makes it increasingly difficult to discern the intent and provenance of statistical methods.

This confusion has serious consequences for the peer review system, as it provides more opportunities for submitters to engage in "p-hacking," cherry-picking algorithms and parameters that return a desired level of significance. Another, perhaps less common, tactic is "steamrolling" reviewers by submitting a novel, opaque algorithm to support a scientific hypothesis. Without reproducible code, evaluating such submissions becomes impossible. Both of these strategies are arrested by reproducible research standards at the publication level.

To test the robustness of a set of results, reviewers should be able to swap in similar methods, but identifying and actually applying an equivalent statistical method is not for the weak of heart. As an example consider gradient boosted trees, a method of building and improving predictive models that involves weak learners (classifiers only slightly better than random guess) using decision trees. Random forests is a popular machine learning algorithm for classification, also decision tree-based. The choice between these two methods is subtle that even experienced data scientists may have to evaluate them empirically but may substantially change model predictions given limited data.

Metadata standards that can support lightweight and heavyweight solutions are well positioned for sustainability and adoption, as are those that provide connections between layers of the analytic stack without a steep learning curve. One example of this which to our knowledge has yet not been implemented is file format and content sanity checks defined by input metadata but implemented at the pipeline level.

Finally there needs to be greater emphasis on translation between embedded and distributed metadata solutions. As discussed, files which support embedded metadata excel as data currency, but may not be ideal for warehousing, querying, or remote access. Conversely, solutions that rely on databases for metadata storage to offer advanced features, whether they be for input metadata, provenance tracking, or workflow execution usually do so at the expense of portability. Systems and standards which provide conduits between these realities are more likely to succeed.

While metadata will always serve as the "who, what, where, why, and how" of data, it is also increasingly the mechanism by which scientific output is made reusable and useful. In our review we have attempted to highlight reproducibility as a vital formal area of metadata research and underscore metadata as an indispensable facet of RCR.

## Acknowledgements

We wish to thank David Clunie, Jian Qin and Farah Zaib Khan for their helpful comments and suggestions.